\documentstyle[12pt]{article}

\begin{document}

\author{M.\ Apostol  \\ 
Department of Theoretical Physics,\\Institute of Atomic Physics,
Magurele-Bucharest MG-6,\\POBox MG-35, Romania\\e-mail: apoma@theor1.ifa.ro}
\title{On the energy spectrum of the $C_{60}^{-}$ fullerene anion}
\date{J.\ Theor.\ Phys.\ {\bf 8} (1995)}
\maketitle

\begin{abstract}
It is shown that the $C_{60}^{-}$ fullerene anion has a hydrogen-like energy
spectrum.
\end{abstract}

The electron affinity of carbon is $1.26\;eV$, while the electron affinity
of the $C_{60}$ fullerene molecule is $2.65\;eV$.\cite{Yang}$^{,}$\cite{Wang}
This large increase in electron affinity can not be accounted by any theory
of chemical bonding.

Typically,\cite{Cioslovski} these theories start with an atomic-like $\pi $%
-type orbital for the added electron, and proceed by taking advantage of the
neighbouring carbon atoms on the surface of the fullerene molecule for
lowering the electron energy. The building blocks of the starting orbital
resemble, more or less, the carbon unoccupied $p_z$-atomic orbital. From
symmetry requirements it is found that the ground-state of the added
electron is triple degenerate and belongs to the $t_{1u}$-representation of
the icosahedral symmetry group of the molecule. These states are thereafter
employed for constructing the energy bands of the alkali ($A$)-doped
fullerides $A_nC_{60}$, yielding a conducting band for the notoriously
insulating $A_4C_{60}$ phase. Small distortions are then invoked, both at
the molecular and the crystalline level, to lift part of degeneracy by
Jahn-Teller effect. However, by such type of approaches, the binding energy
of the added electron is always much smaller than its on-site energy, and,
therefore, the electron affinity of the fullerene molecule could never reach
as large a value as more than twice that of the carbon atom.

The fullerene molecule is practically a spherical molecule of radius $%
R_0=3.5\;\AA $. An electron delocalized over the surface of such a molecule
will induce a polarization charge $z=+e$, where $-e$ is the electron charge.
We may consider in first approximation that this charge is uniformly
distributed over the surface of the molecule. The energy levels of the
electron are, therefore, to be obtained by solving the Schrodinger equation
with the potential energy 
\begin{equation}
\label{one}
\begin{array}{c}
V(r)=-ze/R_0\;\;,\;\;for\;r<R_0\;\;, \\ 
V(r)=-ze/r\;\;,\;\;for\;r>R_0\;\;. 
\end{array}
\end{equation}
We shall use the atomic units $a_H=\hbar ^2/me^2=0.53\;\AA $ (Bohr radius)
and $e^2/a_H=27.2\;eV$, where $m$ is the electron mass and $\hbar $ is
Plack's constant; in addition we put $\hbar =1$, so that $m=e^2=1$. The
solution is of the form $R(\rho )Y_{lm}(\theta ,\varphi )$, where $Y_{lm}$
are the spherical harmonics and $\rho =r/R_0$. It is well-known that the
radial wavefunctions inside the sphere are given by 
\begin{equation}
\label{two}R_{kl}(\rho )=\rho ^l\left( \frac 1\rho \frac \partial {\partial
\rho }\right) ^l\frac{\sin k\rho }\rho \;\;\;, 
\end{equation}
where $l=0,1,2,...$ is the azimuthal quantum number (angular momenta $\hbar
l(l+1)$) and $k^2=2(zR_0-\varepsilon ),\;\varepsilon =-ER_0^2,\;E$ being the
energy, $-zR_0<E<0$. It is also known that these functions are related to
the Bessel functions $J_{l+1/2}(k\rho )$. Outside the sphere the radial
wavefunctions are of the form $F(\rho )e^{-\sqrt{2\varepsilon }\rho }$,
where $F(\rho )$ are polynomials of rank $n$. It is easy to find out that
the spectrum is given by 
\begin{equation}
\label{three}\varepsilon _n=\left( zR_0\right) ^2/2n^2\;\;,\;\;n=1,2,3,... 
\end{equation}
and that there are three types of solutions listed below:

For $l\leq n-1$%
\begin{equation}
\label{four}F_{nl}^{\left( 1\right) }\left( \rho \right)
=\sum_{s=l}^{n-1}\left( -\frac n{2zR_0}\right) ^{n-s-1}\frac{\left(
n-l-1\right) !\left( n+l\right) !}{\left( s-l\right) !\left( l+s+1\right)
!\left( n-s-1\right) !}\rho ^s\;\;\;; 
\end{equation}
For $l\geq n$%
\begin{equation}
\label{five}F_{nl}^{\left( 2\right) }\left( \rho \right)
=\sum_{s=-l-1}^{n-1}\left( \frac n{2zR_0}\right) ^{n-s-1}\frac{\left(
l-s-1\right) !\left( n+l\right) !}{\left( l-n\right) !\left( l+s+1\right)
!\left( n-s-1\right) !}\rho ^s\;\;\;, 
\end{equation}
and 
\begin{equation}
\label{six}F_{nl}^{\left( 3\right) }\left( \rho \right)
=\sum_{s=-l-1}^{-n-1}\left( -\frac n{2zR_0}\right) ^{-n-s-1}\frac{\left(
l-n\right) !\left( n+l+1\right) !}{\left( l-s\right) !\left( l+s+1\right)
!\left( -n-s-1\right) !}\rho ^s\;\;\;. 
\end{equation}
At this point it is worth comparing these results with those corresponding
to the Coulomb potential. In the case of the Coulomb potential, with
wavefunctions finite at $\rho =0$, the electron is pushed further and
further away from the origin with increasing $l$; however, this is not
possible beyond a certain extent, since the Coulomb potential is attractive
enough at the origin, and the electron has to stay there around in order to
take advantage of this attraction.\ Consequently, the azimuthal quantum
number $l$ is bounded from above in the case of the Coulomb potential, as it
is well known. In the present case, the potential is not attractive enough
at the origin, and $l$ may take any positive, integer value. In addition, in
the $\rho >1$ region the second solution of the radial Schrodinger equation
is also alowed, namely that solution which diverges at the origin; it is
given by $\left( 6\right) $ above. It reveals the fact that the electron
tends to stay closer and closer to the surface of the sphere, in order to
diminish its potential energy.

The general solution to our problem for a given $n$ is therefore a
superposition over $l,m$ and the two radial wavefunctions $F_{nl}^{\left(
2,3\right) }$ for $\rho >1$ and $l\geq n$. It is easy to see that the
boundary condition at $\rho =1$, {\it i.e.} the continuity of the
logarithmic derivative, is not satisfied for $l\leq n-1$. Therefore, the
states with low angular momenta are not allowed.\ On the contrary, the
boundary condition is satisfied by a suitable choice of superposition
coefficients for $l\geq n$, due precisely to the presence of the two
independent functions $F_{nl}^{\left( 2,3\right) }$ for $\rho >1$ in this
case.

Summarizing this calculation we may say, therefore, that the energy spectrum
of the electron is 
\begin{equation}
\label{seven}E_n=-\frac{z^2}{2n^2}\;\;,\;\;n=1,2,3,...\;\;\;, 
\end{equation}
and its wavefunctions are (up to a constant) 
\begin{equation}
\label{eight}\Psi _{nlm}=R_{nl}(\rho )Y_{lm}(\theta ,\varphi
)\;\;,\;\;for\;\rho <1\;and\;l\geq n\;\;\;, 
\end{equation}
and 
\begin{equation}
\label{nine}\Psi _{nlm}=\left[ A_{nlm}F_{nl}^{\left( 2\right) }(\rho
)+B_{nlm}F_{nl}^{\left( 3\right) }(\rho )\right] e^{-\sqrt{2\varepsilon _n}%
\rho }Y_{lm}(\theta ,\varphi )\;\;,\;\;for\;\rho >1\;and\;l\geq n\;\;\;, 
\end{equation}
where $R_{nl}$ is given by $\left( 2\right) $ for $k_n^2=2(zR_0-\varepsilon
_n)$ and 
\begin{equation}
\label{ten}
\begin{array}{c}
A_{nlm}= 
\frac{R_{nl}(1)\left[ F_{nl}^{\left( 3\right) ^{\prime }}(1)-\sqrt{%
2\varepsilon _n}F_{nl}^{\left( 3\right) }(1)\right] -R_{nl}^{^{\prime
}}(1)F_{nl}^{\left( 3\right) }(1)}{F_{nl}^{\left( 2\right)
}(1)F_{nl}^{\left( 3\right) ^{\prime }}(1)-F_{nl}^{\left( 3\right)
}(1)F_{nl}^{\left( 2\right) ^{\prime }}(1)}\;e^{\sqrt{2\varepsilon _n}%
}\;\;\;, \\ B_{nlm}=\frac{R_{nl}^{^{\prime }}(1)F_{nl}^{\left( 2\right)
}(1)-R_{nl}(1)\left[ F_{nl}^{\left( 2\right) ^{\prime }}(1)-\sqrt{%
2\varepsilon _n}F_{nl}^{\left( 2\right) }(1)\right] }{F_{nl}^{\left(
2\right) }(1)F_{nl}^{\left( 3\right) ^{\prime }}(1)-F_{nl}^{\left( 3\right)
}(1)F_{nl}^{\left( 2\right) ^{\prime }}(1)}\;e^{\sqrt{2\varepsilon _n}%
}\;\;\;. 
\end{array}
\end{equation}
Apart from the orbital degeneracy the spectrum is otherwise non-degenerate,
as is well known.

The ground-state energy $a$ and the energy $a_1$ of the first excited state
are therefore given by $\left( 7\right) $ as $z^2/2n^2=a$ and $z^2/2\left(
n+1\right) ^2=a_1$; whence we get $n=\left( \sqrt{a/a_1}-1\right) ^{-1}$.
The solution $n$ of this equation should be an integer.\ The ground-state
energy is the electron affinity $a=2.65\;eV$. On the other hand, we know
from absorption experiments\cite{Greaney}$^{,}$\cite{Gasyna} the excitation
energy $1.16\;eV$, whch corresponds to $a_1=1.49\;eV$. Using these data we
get from the above equation $n=2.997$,{\it \ i.e.} $n=3$, and $z=1.32$. One
can check easily that $zR_0<2n^2$ for these values.\ Hence one may conclude
that the energy spectrum of the $C_{60}^{-}$ anion is given by $%
E_n=-z^2/2n^2 $ with $z=1.32$ and $n=3,4,5,...$. The next excited states
have the energies $E_5=-0.95\;eV,\;E_6=-0.66\;eV,\;E_7=-0.48\;eV,\;$etc ($%
E_3=-a=-2.65\;eV,\;$and $E_4=-a_1=-1.49\;eV$). The corresponding absorption
lines have, however, very low intensities.

The orbital degeneracy of the spectrum derived above may indicate that
highly-charged fullerene anions may also exist. However, such a conclusion
is not very likely. Indeed, suppose that we add another electron to $%
C_{60}^{-}$, and neglect for the moment the Coulomb repulsion between the
two electrons.\ We may also assume that each electron induces a charge $z=+1$%
, so that we shall have now a total induced charge $z=+2$. The ground-state
energy in this case will be given by $-2(2^2/2\cdot 3^2)$, which makes $%
-12\;eV$.\ By adding the second electron the fullerene anion gains therefore
an energy $12\;eV-2.65\;eV=9.35\;eV$, which, however, is larger than the
ionization potential $7.6\;eV$ of $C_{60}$.\ This indicates that the
double-charged fullerene anion would not be stable.\ One may lower the
energy gain by including the Coulomb repulsion, which is at most $%
-e^2/2R_0=3.88\;eV$; one can see, nevertheless, that even in this case we
are on the verge of stability with the $C_{60}^{-2}$; and if one uses $%
z=+1.32$ found above, instead of $z=+1$, one obtains a much higher increase
in energy, which definitely renders improbable the formation of the
double-charged fullerene anion. The same conclusion holds also for
highly-charged fullerene anions.

A\ comment is in order here. The value $z=1.32$ obtained above for the
charge induced on the surface of the fullerene molecule, greater that $z=1$
expected for the isolated molecule, may include the polarization effects of
the surrounding environment (solution in the case of Ref.4, solid argon in
the case of Ref.5). This environment, as well as the small irregularities of
the electron density on the surface of the fullerene molecule, may also
split the orbitally degenerate energy levels, leading to a fine structure of
the spectrum; weak absorptions may then be detected, accompanying the main
absorption at $1.16\;eV$. The spectrum may also have an even richer fine
structure, if one considers the coupling of the added electron to the
molecular vibrations.\ It is also worth mentioning in this context that the
rather large values obtained here for the quantum numbers $n$ and $l$
indicate that the non-uniformity of the charge induced on the molecule
surface is small, validating thus our assumption. At the same time, for
these values of the quantum numbers we should expect a rather important
spin-orbit coupling, which is another source (and, probably, the main one)
for the multiplet structure of the spectrum. An interesting type of coupling
will also arise from the magnetic field generated by the induced currents on
the surface of the rotating molecule, which will affect both the orbital
momentum of the electron and its spin; for the former case this is an ${\bf %
L\cdot l}$-type of coupling, where ${\bf L}$ is the angular momentum of the
molecule and ${\bf l}$ is the electron angular momentum. However, the
magnitude of this coupling is extremely small, due, especially, to the large
momentum of inertia of the fullerene molecule.

It might also be worth commenting in this context upon the temptation of
employing the hydrogen-like wavefunctions obtained here for the $C_{60}^{-}$
anion for constructing the energy bands of the alkali-doped fullerides.\
This would not be advisable, since the fullerene molecules in the
alkali-doped fullerides are close enough to one another (the distance
between two neighbouring $C_{60}$'s is $\sim 3\;\AA $) as to disturb
considerably the electron spectrum derived here for a single, isolated
molecule. However, it might be exactly this proximity effect that could
reasonably open another way for constructing the energy bands of these
compounds. Indeed, one may assume that the hybridization of the atomic-like
orbitals of the neighbouring carbon atoms on pairs of adjacent fullerene
molecules are responsible for the electronic structure of these solid-phase
compounds. As it is easy to see, these orbitals were not subject to other
symmetries except those of the (face-centered cubic) crystalline structure.
One may start with the unoccupied carbon orbitals $2p_z,\;3s,\;3p$, etc,
which would lead to non-degenerate bands, predicting thereby that the $%
A_nC_{60}$ fullerides with $n=2,4,6$ are insulating, as pointed out
experimentally. Preliminary calculations show also that $n=6$ seems to be
the highest stable compound in this class.

Finally, we may add that similar calculations as those reported here for $%
C_{60}^{-}$ can also be performed for the $C_{70}^{-}$ anion.\ The
quasi-ellipsoidal shape of the $C_{70}$ molecule\ leads to interesting
effects regarding the degeneracy of the spectrum.

\end{document}